\documentclass[pra, amsmath, amssymb, ,twocolumn,showpacs]{revtex4}
\usepackage{amsfonts}
\usepackage{amsthm}
\usepackage{amstext}
\usepackage{natbib}
\usepackage{hyperref}
\usepackage{graphicx}
\usepackage{psfrag}

\newcommand{\ra}{\rangle}
\newcommand{\la}{\langle}

\DeclareMathOperator{\Tr}{Tr}
\DeclareMathOperator{\ido}{\openone}
\newtheorem{theorem}{Theorem}[section]

\newtheorem{cor}{Corollary}[section]
\newtheorem{con}{Conjecture}[section]
\newtheorem{exam}{Example}[section]
\newtheorem{prop}{Proposition}[section]

\begin{document}
\title{Schmidt balls around the identity}
\author{Lieven Clarisse}
\email{lc181@york.ac.uk}
\affiliation{Dept. of Mathematics, The University of York, Heslington, York YO10 5DD, U.K.}
\begin{abstract}
Robustness measures as introduced by Vidal and Tarrach [PRA, 59, 141-155] quantify the extent to which entangled states remain entangled under mixing. Analogously, we introduce here the Schmidt robustness and the random Schmidt robustness. The latter notion is closely related to the construction of Schmidt balls around the identity. We analyse the situation for pure states and provide non-trivial upper and lower bounds. Upper bounds to the random Schmidt-2 robustness allow us to construct a particularly simple distillability criterion. We present two conjectures, the first one is related to the radius of inner balls around the identity in the convex set of Schmidt number $n$-states. We also conjecture a class of optimal Schmidt witnesses for pure states.
\end{abstract}

\maketitle

\section{Introduction}
There are two obvious ways of quantifying \cite{Horodecki01, PV05} entanglement, operational and geometrical. Operational entanglement measures include the distillable entanglement $E_D$ \cite{BDSW96, Rains99} and the entanglement cost $E_C$ \cite{BDSW96, HHT00}, and are directly related to the physical operations of extracting entanglement from a state and constructing the state back from maximally entangled singlets. Geometrical measures can loosely be described as those quantifying the distance from a state to the set of separable states. Examples include the relative entropy of entanglement \cite{VPRK97,VP98}, the negativity \cite{VW01}, the GME \cite{WG03}, the best separable approximation \cite{KL00} and the robustness measures \cite{VT99, Steiner03, HN03}.
Many of these geometrical measures can be cast directly into the language of entanglement witnesses \cite{Brandao05}, as we will illustrate later for the random robustness.

Entanglement witnesses were originally introduced as a way of `detecting' entanglement. The basic idea is that the set of separable states is convex. The following theorem gives a geometrical characterization of the problem of determining whether a state $\rho\in D$ is contained in a certain compact and convex subset $S\subset D$:
\begin{theorem}[\cite{HHH96, BCHHKLS01}]
\label{separation}
Let $S\subset D$ be a convex compact set of states and $\rho\in D$. If $\rho\notin S$, then there exists a Hermitian operator $W$ such that $\Tr{(W\rho)}<0$ and $\Tr{(W\sigma)}\geq 0$ for all $\sigma \in S$.
\end{theorem}
This theorem is an immediate consequence of basic theorems in functional analysis \cite{Rockafellar70, Lax02}. Namely, the Hahn-Banach theorem states that a convex set and a point lying outside it can be separated by a hyperplane $W$, and the Riesz-Frechet representation theorem then characterizes such hyperplanes. The hyperplanes $W$ are commonly called witnesses \cite{Terhal01} (they witness states outside $S$). Clearly, we do not need all possible witnesses to characterize $S$, it is enough to consider those witnesses tangent to $S$ (a witness $W$ is tangent to $S$ if there exists a $\rho \in S$ such that $\Tr(W\rho)=0$). When $S$ is the set of the separable states $W$ is called an entanglement witness. Here $W$ is positive on separable states and negative on at least one entangled state. 
From an entanglement witness $W$ one can construct a geometrical entanglement measure $E_M$ as follows \cite{Brandao05}
\begin{align}
E_M(\rho)=\max_{W\in M}[0, - \Tr (\rho W )],
\end{align}
where $M$ is some compact subset of the set of entanglement witnesses. 

In this work we will focus on robustness measures. Let us therefore recall their definition. 
We define the $K$-\emph{robustness} of a state $\rho$, $R_k(\rho)\geq 0$ as the minimal value of $R$ such that 
\begin{align}
\frac{1}{1+R}(\rho+R\rho_k)
\end{align}
is separable, for some state $\rho_k \in K$. With this basic definition the \emph{robustness} $R_s(\rho)$ of a state $\rho$  as introduced by Vidal and Tarrach \cite{VT99} equals the $S$-robustness, with $S$ the set of separable states.  The \emph{random robustness} $R_r(\rho)$  is defined as the $\ido$-robustness ($K=\{\ido\}$)\footnote{For simplicity we have chosen not to normalise $\ido$ in the definition, in contrast to the original definition of random robustness in \cite{VT99}.}. Here $\ido$ is the identity operator (the unnormalised totally mixed state). Finally the \emph{generalised robustness} \cite{Steiner03, HN03} is defined as the $D$-robustness, with $D$ the set of all normalised states. Thus robustness measures measure how much mixing is required before a state becomes separable. It is also clear that $R_g \leq R_s \leq R_r$. 

Both the robustness and the generalised robustness are entanglement monotones. Recently it has emerged \cite{VV03, Brandao05b} that the generalised robustness has a very nice operational meaning as the maximum percentuel increase an entangled state can provide in the fidelity of teleportation of another state. It is easy to see from the definition that the random robustness is proportional to the so-called witnessed entanglement \cite{BV04b}, defined as $\max_{W\in M}[0, - \Tr (\rho W )]$ with $\Tr(W)=1$. Also note that for a given dimension, the maximum random robustness over all entangled states gives rise to a lower bound on the volume of separable states \cite{VT99}. The following theorem gives exact values for the robustness and the random robustness for pure states.

\begin{theorem}[\cite{VT99, Steiner03, HN03}]
\label{rrth}
Let $|\psi\ra=\sum_i a_i |ii\ra$ be a pure bipartite state with ordered Schmidt coefficients $a_i\geq a_{i+1}$. 
The robustness $R_s$ of $|\psi\ra$ is given by $R_s(\psi)=\sum_{i\neq j}a_i a_j =(\sum_i a_i)^2-1$ and equals its generalised robustness.
A separable state that washes out the entanglement in $\psi$  most quickly is given by 
\begin{align}
\rho_s=\rho_g=\frac{1}{R_s}\sum_{i \neq j} a_i a_j |ij \ra\la ij|.
\end{align}
The random robustness of $\psi$ is given by $R_r(\psi)=a_1a_2$.
\end{theorem}
An alternative proof of this theorem can be obtained as a corollary of our results (see Section~II).

\section{Robustness measures of Schmidt number}
In this section we will extend the notion of generalised robustness to generalised Schmidt robustness.

Let us first recall the Schmidt decomposition \cite{Schmidt06, NC00} of a pure state $|\psi\ra\in {\cal H}={\cal H}_A \otimes {\cal H}_B$. It can be shown that there always exist orthonormal basis states $|i_A\ra$ and $|i_B\ra$ for ${\cal H}_A$ and ${\cal H}_B$ respectively such that $|\psi\ra=\sum_{i=1}^k a_i |i_A\ra|i_B\ra$, where the $k$ numbers $a_i$ are non-negative real numbers satisfying $\sum_i a^2_i=1$, known as the Schmidt coefficients. We call $k$ the Schmidt rank of $\psi$. For the sequel we make the convention that we order the Schmidt coefficients as $a_i\geq a_{i+1}$ for all $i$.

This definition can be extended to mixed states \cite{TH00} as follows. A bipartite mixed state $\rho$ acting on ${\cal H}={\cal H}_A \otimes {\cal H}_B$  is said to have Schmidt number $n$ if there exists a decomposition of $\rho=\sum_i p_i |\psi_i\ra \la\psi_i |$ with all vectors $\{|\psi_i\ra\}$ having Schmidt rank at most $n$, and there exists no such decomposition with all vectors having a Schmidt rank $n-1$ or lower.
This definition coincides with the previous one in the case where $\rho$ is a pure state. 
Separable states have Schmidt number one, and entangled states have Schmidt number larger than one.
It is convenient to denote the set of all density operators as $D$ and the set of density operators that have Schmidt number $n$ or less by $S_n$. Sets of increasing Schmidt number are embedded into each other as $S_1 \subset S_2  \subset \ldots \subset S_d=D$. The subsets $S_i$ are all convex and compact by construction.

We call a hermitian operator $W$ a Schmidt witness of class $n$ (for short $n$-SW) \cite{SBL00, HBLS04} if and only if
\begin{enumerate}
\item $\Tr(W\sigma)\geq 0$ for all $\sigma \in S_{n-1}$,
\item There exists a $\rho \in S_n$ such that $\Tr(W\rho) < 0$,
\end{enumerate}
The existence of a $n$-SW for a Schmidt number $n$ state follows just from Theorem~\ref{separation}.

\begin{exam}[\cite{TH00,SBL00}]
Let $P_+$ be a maximally entangled state acting on a Hilbert space ${\cal H}\cong \mathbb{C}^d \otimes \mathbb{C}^d$ and consider the unnormalized isotropic states defined by
\begin{align}
\rho_\beta=\ido+\beta P_{+}, \qquad \mbox{with} \qquad -1\leq  \beta \leq  \infty.
\end{align}
The isotropic state $\rho_\beta$ has Schmidt number $n$ if and only if
\begin{align}
\frac{d((n-1)d-1)}{d-(n-1)} <  \beta \leq \frac{d(nd-1)}{d-n}.
\end{align}
The operators 
\begin{align}
\label{canwit}
W_n=\ido - d/(n-1) P_+,
\end{align}
are $n$-SW and detect in an optimal way the Schmidt number of the states $\rho_\beta$.
\label{impex}
\end{exam}

In what follows we assume that all states act on a Hilbert space ${\cal H}\cong \mathbb{C}^d \otimes \mathbb{C}^d$.

\subsection{Generalised Schmidt robustness}
Analogously to the generalised robustness, the generalised Schmidt-$n$ robustness of a state $\rho$, $R_{gn}(\rho)$ is defined as the minimal value of $R$ such that 
\begin{align}
\frac{1}{1+R}(\rho+R\rho_{gn})
\end{align}
has Schmidt number smaller or equal than $n$, for some $\rho_{gn}$.

Let us now analyse the generalised Schmidt robustness for pure states. 
In Theorem~\ref{rrth} we have seen that for a pure state $|\psi\ra=\sum_i a_i |ii\ra$ the state
\begin{align}
\rho_g=\frac{1}{R_g}\sum_{i \neq j} a_i a_j |ij \ra\la ij|,
\end{align}
erases most quickly the entanglement present in $|\psi\ra$. For the maximally entangled state $|\psi_+\ra=\sum_i \frac{1}{\sqrt{d}}|ii\ra$ this state is given by $\rho_g=\frac{1}{d^2-d} (\ido-Z)$, with $Z=\sum_i|ii\ra\la ii|$. It is very plausible that this same state will also erase Schmidt number in an optimal way. We were able to prove this for the maximally entangled state:

\begin{theorem}[Generalised Schmidt robustness of the maximally entangled state]
\label{gsrmes}
The state defined by 
\begin{align}
\rho(\beta)=\frac{\beta \rho_g + P_+}{1+\beta} \quad \text{where} \quad  \rho_g=\frac{\ido-Z}{d^2-d}
\label{tooproof}
\end{align}
has Schmidt number $n$ for 
\begin{align}
\frac{d-n}{n} \leq \beta < \frac{d-n+1}{n-1}. 
\end{align}
The generalised Schmidt-$n$ robustness of the maximally entangled state $P_+$ is given by $R_{gn}(P_+)=\frac{d-n}{n}$.  
\end{theorem}
\begin{proof}
Let $S(\beta)$ be the Schmidt number of $\rho(\beta)$ and let $\beta_n=\frac{d-n}{n}$. We first show that $S(\beta_n)\leq n$.

(i) We will give an explicit decomposition of the state $\rho(\beta_n)$ in terms of Schmidt rank $n$ states. Equivalently, we show how one can construct this state locally with the aid of Schmidt rank $n$ states. In what follows we will often omit normalisation. Let us take a maximally entangled $n$-level state 
\begin{align}
|\psi_S\ra=\frac{1}{n}\sum_{i\in S}^n|ii\ra,
\end{align}
where $S=\{i_1,\ldots,i_n \}$ is a subset of $\{1,\ldots,d\}$. We can construct such a state in $\binom{d}{n}$ possible ways, and clearly all these states have Schmidt number $n$. Now let us mix with equal weight the corresponding states $|\psi_S\ra\la\psi_S|$. Then for every $i$ and $j$ we will have
$\binom{d-1}{n-1}$ terms of the form $|ii\ra \la ii|$ and $\binom{d-2}{n-2}$ terms of the form $|ii\ra \la jj|$, $i\neq j$. Thus the resulting state will be proportional to 
\begin{align}
(d-1)Z+(n-1)\sum_{i\neq j} |ii\ra \la jj|.
\end{align}
Therefore we have proven that the state (see \cite{Clarisse04} for n=2)
\begin{align}
K=Z +\frac{d(n-1)}{d-n}P_+
\end{align}
has Schmidt number at most $n$. It turns out that we can transform this state in the form (\ref{tooproof}) by applying a certain partial twirl operation. Consider the following twirl operation \cite{HH97,Rains98,VW01}
\begin{align}
\int dU (U\otimes U^*) \rho (U\otimes U^*)^\dagger,
\end{align}
which maps any state $\rho$ into one of the form $\ido + \alpha P_+$ (an isotropic state). Here $dU$ is the uniform probability distribution on the unitary group $U(d)$. Note that the twirl can be implemented locally, both parties need to implement only one random unitary. Remarkably, it has been shown \cite{DCLB99} \footnote{In Ref.\ \cite{DCLB99} the finite decomposition of the $U\otimes U$ twirl was given, but a  similar decomposition can be deduced from it for the $U\otimes U^*$ twirl.} that the integral can be written as a finite sum (for qubits this was first shown in Ref.\ \cite{BDSW96}). This will allow us to perform a partial twirl, just by considering a part of this finite sum. 

In the first step we apply the unitary transformation $(T\otimes T^*) K (T\otimes T^*)^\dagger$, with
\begin{align}
T=\frac{1}{\sqrt{d}} \sum_{j,k=0}^{d-1} e^{\frac{i 2\pi jk}{d}}|j\ra\la k|,
\end{align}
which is just the quantum fourier transform. Since $T$ is unitary, it acts as the identity on $P_+$ \cite{HH97}, while it acts on $Z$ as
\begin{align}
(T\otimes T^*) Z (T\otimes T^*)^\dagger = \nonumber\\
 \frac{1}{d^2}\sum_a \sum_{j,k,s,t} e^{\frac{i 2\pi (j-t+k-s)a}{d}}|j\ra\la t| \otimes |s\ra\la k|.
\end{align}
The terms of the form $|jj\ra\la kk|$ for $s=j$ and $t=k$ will give a contribution $P_+$, while the $|ij\ra\la ij|$ for $i\neq j$ (for $j=t$ and $k=s$) will give a contribution of $(\ido-Z)/d$. So that 
\begin{align}
K'=d (T\otimes T^*) K (T\otimes T^*)^\dagger= \ido -Z + \frac{d(d-1)n}{d-n}P_+ + L,
\end{align}
where $L$ are terms not of the form  $|ii\ra\la jj|$ or $|ij\ra\la ij|$. Now these contributions can be easily removed by repetitive application of the following mixing procedure 
\begin{align}
K''=\frac{1}{2} U\otimes U^* K  (U\otimes U^*)^\dagger + \frac{1}{2}K'.
\end{align}
First $U$ is chosen to act as $U|k\ra=e^{i\pi\delta_{kl}}|k\ra$ for every $l=0,\ldots ,d-1$. This defines $d$ mixing procedures. Next $U$ is taken to act as $U|k\ra=e^{i\pi\delta_{kl}/2|k\ra}$ (another $d$ mixing procedures). One can readily check \cite{DCLB99} that these operations do not affect terms 
$|ii\ra\la jj|$ or $|ij\ra\la ij|$ but cancel out $L$ completely.

 Thus $S(\beta_n) \leq n$. Now for $\beta_n\leq \beta < \beta_{n-1}$, the state $\rho(\beta)$ is convex combination of $\rho(\beta_n)$ and $\rho(\beta_{n-1})$ and therefore $S(\beta) \leq n$.

(ii) For the second part, we generalise the trick introduced in Ref.\ \cite{HN03}. For any state $\sigma$, suppose $t$ is a positive number such that $P_+ +t \sigma$ has Schmidt number $n$. The operators $W_n=\ido - d/n P_+$ witness Schmidt number $n+1$ (see Example~\ref{impex}), so that we have
\begin{align}
0 & \leq \Tr{ [ (\ido-d/nP_+)(P_+ + t \sigma)] }\nonumber\\ 
  & = 1+t-d/n\Tr{[P_+]}-d/n\Tr{[P_+ \sigma]} \nonumber\\ 
  & \leq - \frac{d-n}{n} + t.
\end{align}
since $\Tr{P_+ \sigma}\geq 0$. Thus $t\geq \beta_n$, and for $\sigma=\rho_g$ it follows that $S(\beta)\geq n$ for $\beta_n\leq \beta < \beta_{n-1}$.

For general $\sigma$, it follows that $R_{gn}\geq \beta_n$, but $\rho(\beta_n)=P_+ + \beta_n \rho_g$ has Schmidt number $n$, so that $R_{gn}(P_+)=\beta_n$.
\end{proof}

Note that the states (\ref{tooproof}) constitute one of the very few examples of non-trivial one parameter states for which the Schmidt number is known. To our knowledge, the only other example is that of the isotropic states, Example~\ref{impex}. This theorem allows us to present non-trivial bounds to the generalised Schmidt robustness of arbitrary pure states.\vspace{0.001cm}

\begin{cor}[Lower and upper bounds for the generalised Schmidt robustness]
\label{corgrob}
The generalised Schmidt-$n$ robustness $R_{gn}$ of a pure state $|\psi\ra=\sum_i a_i |ii\ra$ satisfies
\begin{align}
 \frac{1}{n}(\sum_i a_i)^2-1 \leq R_{gn}  \leq R_g \frac{d-n}{(d-1)n},
\end{align}
with $R_g=\sum_{i\neq j}a_i a_j =(\sum_i a_i)^2-1$ the generalised robustness of $\psi$.
\end{cor}

\begin{proof}
In Theorem~\ref{gsrmes} we have seen that $\rho=\frac{d-n}{n} \frac{1}{d^2-d}(\ido-Z) + P_+$ has Schmidt number $n$. Performing the filtering operation $(A\otimes A) \rho (A\otimes A)^\dagger$ we hence obtain a state with Schmidt number at most $n$ (this is because local filtering cannot increase the Schmidt number of a state \cite{TH00}). If we take $A=\sum_k \sqrt{a_k} |k\ra \la k|$ we obtain that
\begin{align}
|\psi \ra\la \psi| + R_g \frac{d-n}{(d-1)n}\rho_g
\end{align}
has Schmidt number $n$. Here $\rho_g=\frac{1}{R_g}\sum_{i \neq j} a_i a_j |ij \ra\la ij|$ as before. This gives the upper bound. The lower bound can be readily proven using exactly the same trick as in part (ii) of Theorem~\ref{gsrmes}.
\end{proof}
The lower and upper bound only coincide in the case $\psi$ is the maximally entangled state or when $n=1$. Note that the lower bound can be negative. The upper bound depends on the dimension of the Hilbert space in which the state is embedded, and hence will in general not match the value of the generalised Schmidt robustness.

\subsection{Random Schmidt robustness}
We define the random Schmidt-$n$ robustness of a state $\rho$, $R_{rn}(\rho)$ as the minimum value of $R$ such that
\begin{align}
\frac{1}{1+R}(\rho+R\ido)
\end{align}
has Schmidt number $n$.

As we have seen from Example~\ref{impex} for $\rho=P_+$ we have $R_{rn}=(d-n)/[d(nd-1)]$. For general pure states a (weak) upper bound to the random Schmidt robustness can be obtained as follows. We know that
\begin{align}
\Gamma_n=(d-n)\ido +(nd-1)dP_+
\end{align}
has Schmidt number $n$. Local filtering $(A\otimes A) \Gamma_n (A\otimes A)^\dagger$ cannot increase the Schmidt number. So that with $A=\sum_k \sqrt{a_k} |k\ra \la k|$ we obtain
\begin{align}
\Gamma'_n=(d-n)\rho_A \otimes \rho_A +(nd-1)|\psi\ra\la\psi|,
\end{align}
with $\rho_A$ the reduced density operator of $|\psi\ra\la\psi|$, and because  $\ido-\rho_A\otimes \rho_A$ is a separable state we get as an upper bound for the random Schmidt robustness $R_{rn}(\psi) \leq (d-n)/(nd-1)$. The following theorem presents a non-trivial upper bound.

\begin{theorem}[Upper bound to the random Schmidt robustness]
The random Schmidt-$n$ robustness $R_{rn}$ of a pure state $|\psi\ra=\sum_i a_i |ii\ra$ satisfies
\begin{align}
R_{rn}  \leq  \frac{R_r(d-n)}{dn-1},
\end{align}
with $R_r=a_1a_2$ the random robustness of $\psi$.
\end{theorem}
\begin{proof}
In this proof, we work again with unnormalised states. Note that, if we add two unnormalised Schmidt number $n$ states together we end up with a Schmidt number $n$ state. Equivalently, this can be shown by mixing the normalised states with different weights.

Analogously to the construction in Theorem~\ref{gsrmes}, let us take a maximally entangled $n$-level state 
\begin{align}
|\psi\ra_{S}=\frac{1}{n}\sum_{i\in S}^n a_i |ii\ra,
\end{align}
where $S=\{i_1,\ldots,i_n \}$ is a subset of $\{1,\ldots,d\}$. Again, we can construct such a state in $\binom{d}{n}$ possible ways, and mixing the corresponding (unnormalised) states $|\psi_S\ra\la\psi_S|$ together, we end up with a state proportional to
\begin{align}
\sum_i a_i^2|ii\ra \la ii| +\frac{(n-1)}{d-n} |\psi\ra\la\psi|,
\end{align}
which has at most Schmidt number $n$. Now, in Corollary~\ref{corgrob} we have seen that the state
\begin{align}
\frac{(d-1)n}{d-n} |\psi \ra\la \psi| +  \sum_{i \neq j} a_i a_j |ij \ra\la ij|
\end{align}
has Schmidt number at most $n$. Adding these two states together we find that 
\begin{align}
\frac{dn-1}{d-n} |\psi \ra\la \psi| +  \sum_{i, j} a_i a_j |ij \ra\la ij|
\end{align}
has Schmidt number no more than $n$. Mixing this state with the separable state $\sum_{i,j}(a_1a_2-a_ja_j)|ij\ra\la ij|$, we obtain finally 
\begin{align}
\frac{dn-1}{(d-n)a_1a_2} |\psi \ra\la \psi| +   \ido.
\end{align}
\end{proof}

Upper bounds to the random Schmidt-2 robustness are particularly useful, since they give rise to the following distillability criterion:

\begin{prop}[Distillability criterion]
Let $\rho$ be an arbitrary bipartite state, such that $\rho^{T_B}$ (the partial transposition \cite{Peres96} of $\rho$) has negative eigenvalues. Let $|\psi\ra$ be the eigenvector corresponding to a negative eigenvalue $\lambda<0$ and let $ \tilde{R}_{r2}(\psi)$ be an upper bound to its random Schmidt-2 robustness. Then $\rho$ is distillable if 
\begin{align}
\lambda < - \tilde{R}_{r2}(\psi).
\end{align}
\end{prop}
\begin{proof}
From the definition, we have that
\begin{align}
W=|\psi\ra\la\psi|+\tilde{R}_{r2}(\psi)\ido
\end{align}
has Schmidt number two. Now if \cite{Clarisse04}  $\Tr (W \rho^{T_B})<0$ then $\rho$ is distillable. This can we rewritten as
\begin{align}
\la\psi|\rho^{T_B}|\psi\ra + \tilde{R}_{r2}(\psi)= \lambda + \tilde{R}_{r2}(\psi) <0.
\end{align}
\end{proof}

This proposition provides an important reason to find an exact analytical formula for $R_{r2}(\psi)$. 
Note that this distillability criterion only depends on the minimum eigenvalue and the corresponding eigenvector. It is easy to see that it is independent of another simple distillability criterion, the reduction criterion. The reduction criterion \cite{HH97, CAG97} says that when a state $\rho$ satisfies $\openone\otimes \rho_B-\rho \not \geq 0$ or $\rho_A\otimes\openone -\rho\not \geq 0$ then $\rho$ distillable. Consider the following rather extreme example:
\begin{align}
\rho=\frac{1}{16}\left[ \begin{array}{rrrrrrrrr}
  1 & 0  & 0 & 0  & 0 & 0 & 0 & 0 & 0 \\
  0 & 2  & 0 & -1 & 0 & 0 & 0 & 0 & 0 \\
  0 & 0  & 2 & 0  & 0 & 0 & 2 & 0 & 0 \\
  0 & -1 & 0 & 2  & 0 & 0 & 0 & 0 & 0 \\
  0 & 0  & 0 & 0  & 1 & 0 & 0 & 0 & 0 \\
  0 & 0  & 0 & 0  & 0 & 2 & 0 & 2 & 0 \\
  0 & 0  & 2 & 0  & 0 & 0 & 2 & 0 & 0 \\
  0 & 0  & 0 & 0  & 0 & 2 & 0 & 2 & 0 \\
  0 & 0  & 0 & 0  & 0 & 0 & 0 & 0 & 2
\end{array} \right].
\end{align}
It is easy to check that the reduction criterion is not useful here. The partial transposition $\rho^{T_B}$
has an eigenvector $|\psi\ra=\frac{1}{\sqrt{3}}( |00\ra+|11\ra-|22\ra)$ and corresponding eigenvalue $-\frac{1}{8}$. Since $-{R}_{r2}(\psi)=-\frac{1}{15}>-\frac{1}{8}$, $\rho$ is distillable. 

An unconditional distillability criterion, which does only depend on the minimum eigenvalue of $\rho^{T_B}$ would follow from the following conjecture. Note that it is a generalisation of the fact that  $\ido +2\rho$ is separable for all normalised states $\rho$ \cite{VT99}.

\begin{con}[Schmidt balls around the identity]
\label{mainconject}
Consider the unnormalized mixture
\begin{align}
\rho_\beta=\ido+\beta \rho, \qquad \mbox{with} \qquad -1\leq  \beta \leq  \infty,
\label{rbbb}
\end{align}
with $\rho$ an arbitrary normalized state acting on  $\mathbb{C}^d \otimes \mathbb{C}^d$. 
The state $\rho_\beta$ has Schmidt number at most $k$ for \emph{all} $d$ and $\rho$ when
\begin{align}
\beta \leq 2(2k^2-1).
\label{vvalue}
\end{align}
\end{con}

Pending on the proof of this conjecture, we have that $\rho$ is distillable whenever $\rho^{T_B}$ has an eigenvalue $\lambda \leq -1/14$. The conjectured value (\ref{vvalue}) can be obtained as follows.
Starting from the identity, it is natural to assume that we can go fastest to a higher Schmidt number by mixing with some maximally entangled state. Taking $\rho=P^{d'}_+$ (the maximally entangled $d'$-level state) we have that $\beta\leq d'(kd'-1)/(d'-k)$, which reaches its maximum for $d'=\sqrt{k^2-1}+k$ or for integer $d'=2k$. Substituting this expression in the expression for $\beta$ gives the upper bound from the conjecture. Thus it looks like, starting from the identity, we can get most quickly to a Schmidt number $k+1$ state by mixing with the maximally entangled state in $2k \otimes 2k$.

Lower bounds on the random Schmidt robustness of a pure state $\psi$ can obtained from any Schmidt witness that detects $\psi$. Indeed, suppose $W_{n+1}$ is a normalised Schmidt number $n+1$ witness such that  $\Tr(|\psi\ra\la\psi| W_{n+1})=-\alpha$, with $\alpha>0$. Then it is easy to see that 
$R_{rn}(\psi) \geq \frac{\alpha}{d^2}$. It follows that
\begin{align}
d^2R_{rn}(\psi)= \min_{Tr W_{n+1}=1} - \Tr(|\psi\ra\la\psi| W_{n+1}).
\end{align}
Consider the Schmidt witnesses $W_{n+1}=\ido - d/n P_+$ from Example~\ref{impex}. Performing the filtering operation $(A\otimes B) W_{n+1} (A\otimes B)^\dagger$ we obtain again a Schmidt number $n+1$ witness. Let us consider the particular case where $A=B=\sum_k \sqrt{a_k} |k\ra \la k|$ diagonal and such that $\sum_i a_i^2=1$; we obtain
\begin{align}
\label{clclcl}
W_{n+1}=\sum_{ij} a_i a_j |ij \rangle \langle ij | - \frac{1}{n}|\psi\ra\la\psi|.
\end{align}
The trace is given by $(\sum_i a_i)^2-1/n$, and therefore the value of the normalised $W_{n+1}$ on an arbitrary pure state $|\psi\rangle=\sum_i b_i|ii\ra$ is 
\begin{align}
\frac{n\sum_i a_i^2 b_i^2-(\sum_i a_i b_i)^2}{n(\sum_i a_i)^2-1}.
\end{align}
For arbitrary pure states, this class of witnesses gives the highest value on lower bounds of $R_{rn}$ we have found, therefore we conjecture 

\begin{con}
The random Schmidt robustness $R_{rn}$ of a pure state $|\psi\ra=\sum_i b_i |ii\ra$ is given by
\begin{align}
\label{dddd}
R_{rn}(\psi)= d^2 \max_{a_i } -\frac{n  \sum_i a_i^2 b_i^2-(\sum_i a_i b_i)^2}{n(\sum_i a_i)^2-1},
\end{align}
with $\sum_i a_i^2=1$.
\label{swcc}
\end{con}

\begin{figure}
\psfrag{R}{$R_{r2}$}
\psfrag{a}{$a_1^2$}
\center{\includegraphics[width=7.5cm]{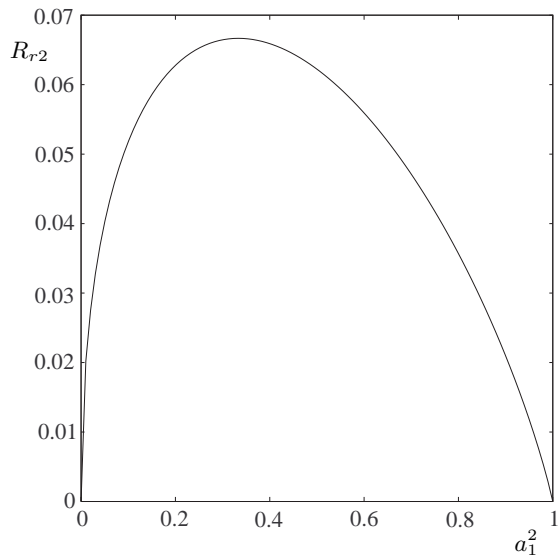}}
\caption{Lower bounds on $R_{r2}(\psi)$ in function of $a^2_1$, where $|\psi\ra = a_1|00\ra + a_2(|11\ra+|22\ra)$. The graph was obtained numerically using the witnesses from Conjecture~\ref{swcc}. The best fit we found was $R_{r2}(\psi)=0.15 a^{0.85}_1 (1-a^2_1)^{0.85}$.}
\label{figsw}
\end{figure}

A first step in proving this conjecture would be to show that the class of witnesses $(A\otimes B) W_{n+1} (A\otimes B)^\dagger$ with $A,B$ arbitrary matrices, is no more powerful than the class of witnesses $(A\otimes A) W_{n+1} (A\otimes A)^\dagger$ , with $A$ diagonal. We have numerically verified this for $n=2,3$ and $d=3,4$. Another open problem is the  evaluation of the maximisation in Equation~(\ref{dddd}). Ideally we would like to have an expression only in terms of the coefficients $b_i$.
In Figure~\ref{figsw} we have plotted the numerical maximization of Equation~(\ref{dddd}) for a particular set of pure states (see caption).

In conclusion, we have presented strong upper and lower bounds for the generalised and random Schmidt robustness for pure states. The problem of finding exact values is very hard, as in the end, one has to come up with an explicit convex decomposition in Schmidt rank $n$ states on the one hand, and on the other with a construction of optimal Schmidt witnesses. We hope that our results may stimulate further work, especially in proving or disproving Conjecture~\ref{mainconject}.

\begin{acknowledgments}
This work was supported by a WW Smith Scholarship. I would like to thank Anthony Sudbery for numerous discussions on this work, Pawe{\l} Horodecki for a lengthy discussion and Florian Hulpke for a discussion on Ref.\ \cite{HBLS04}. I am also grateful to Sibasish Ghosh, Samuel Braunstein, Simone Severini and William Hall for discussions. I am very grateful to Christine Aronsen Storebo for support.
\end{acknowledgments}

\bibliographystyle{apsrevl}
\bibliography{ent}

\end{document}